# Water transfer and crack regimes in nano-colloidal gels

J. Thiery, S. Rodts, E. Keita, X. Chateau, P. Faure, D. Courtier-Murias, T. E. Kodger*, P. Coussot

Laboratoire Navier (ENPC-CNRS-IFSTTAR), Université Paris-Est

* Experimental Soft Matter Group, Harvard University

**Abstract**: Direct observations of the surface and shape of model nano-colloidal gels associated with measurements of the spatial distribution of water content during drying show that air starts to significantly penetrate the sample when the material stops shrinking. We show that whether the material fractures or not during desiccation, as air penetrates the porous body, the water saturation decreases but remains almost homogeneous throughout the sample. This air-invasion is at the origin of another type of fracture due to capillary effects; these results provide a new insight in the liquid dynamics at the nano-scale.

## I. INTRODUCTION

The evaporation of solvent from colloidal suspensions, gels or pastes often leads to shrinking, wrapping, and eventually cracking [1] which impacts the material. Cracking during drying raises tremendous technological and scientific concerns as it can drastically impair the structural properties of a concrete [2], the integrity of ceramic films [3], soil or clays [4], the functionalities of cosmetics, paints or coatings [5].

Among the different physical mechanisms [6] suggested to explain why a crack nucleates during drying the most common model relies on the framework of elastic fracture mechanics [7]. This tensile failure theory considers the propagation of cracks triggered once the near tip stress state which intensifies as desiccation proceeds if the material remains bound to its substrate reaches the failure criterion for brittle materials [4, 8-9]. By contrast, the opening of cracks during drying is considered to stem from large capillary stresses resulting from the penetration of irregular air-liquid interface into the material [6, 10]. However, even though the examination of fracture patterns as a function of experimental conditions is considerable [11], no study bears direct witness of these effects at local length scale. Additionally, despite drying regimes are already accurately detailed with the spatial distribution of the liquid in time [12-13], the role of liquid dynamics on fracturing appear to be somewhat neglected so far notwithstanding that the tensile failure theory implicitly assumes the sample to remain saturated with a rather homogeneous distribution of water, whereas, its counterpart brings up the idea of a partially saturated medium, at least locally. Therefore, measurements of the spatial distribution with time of water during drying would greatly enhance the current understanding of crack nucleation.

In this paper we follow both the surface and internal characteristics, specifically the spatial distribution of water, of a drying model nano-colloidal gel with the help of proton NMR profiling (a 1D use of Magnetic Resonance Imaging) measurements with micrometre spatial resolution. We show that whether it develops widely open fractures or not during the process, the material remains saturated as it shrinks, and then starts to desaturate almost homogeneously, which induces thin fractures that do not span the gel. These results provide a general view of water transport during drying and cracking, and a new insight in the thin liquid film dynamics in nanopores.

## II. MATERIALS AND METHODS

We use aqueous dispersions of hydrophilic nano-sized silica particles (*r*=6 nm radius) obtained from Sigma Aldrich (Ludox HS-40) in which 0.5mol/L of Sodium chloride is added to trigger aggregation. The initial stable suspension of particles has a solid volume fraction of 23%. Since according to the DLVO theory the electrical double layer thickness around a single particle is governed by the concentration of the surface charge counterions in liquid phase of the suspension, the higher the salt concentration in this electrolyte, the higher the aggregation rate of colloids. Therefore, in order to obtain a fairly homogeneous final gel structure i.e avoiding substantial structural heterogeneity possibly caused by local gradients of rate of aggregation — sedimentation not applying here— we dissolved the quantity of salt needed into the additional necessary water (thus mixing two homogeneous phases together) to work at initial volume fraction of particles ($\phi_0$) ranging from 5% to 20%.

The mixture processed being firstly liquid, we poured it into glass Petri dishes of 90 mm diameter, therefore making the initial sample thickness ($h_0$) either 5 or 10 mm, and further sealed the recipient and leave the mixture at rest for gelation time. The material structure in time is followed by rheometrical tests imposing oscillations at very low strain ($3.10^{-4}$) (frequency: 1 Hz) that do not interfere with gelation. The storage modulus ($G'$) increases by several orders of magnitude and finally, typically after one or two hours, reaches a plateau that we associate with the steady gel structure. The values of $G'$ at the plateau were typically in the range $10^5$-$10^6$ Pa, which means that we are dealing with initially rigid samples.

In order to avoid a possible effect on drying characteristics of the presence of salt in solution we removed 97% of the salt initially incorporated through a dialyse protocol. It consists to cyclically pour de-ionized water on top of the sample, then waiting the time needed for the sodium chloride to homogenize throughout this new two-phase system, and finally carefully removed the liquid. Each dialyze enabled us to remove roughly half the quantity previously contained in the gel within a very small amount of time which actually depends on the sample volume. The amount of NaCl extracted after each dialyze was evaluated performing conductivity measurements on dialysates. The remaining very low concentration of salt is expected to negligibly affect water transfer processes, and the diffusion of salt in water being much more rapid than advection due to liquid motion, no significant accumulation of salt crystals should play a role in the fracture processes.

In order to look at the impact of dialyze on gel strength we managed to dialyze the sample during rheometrical tests thanks to a special set up allowing a water bath around the geometry. We observed a drop of about 25% of the $G'$ value when water is first put in contact with the sample, likely due to a structure modification of the gel generated by the osmotic pressure of de-ionized water in the bath, but subsequent dialyzes did not further affect $G'$. This means that dialyze negligibly affects the strength of the structure which relies on strong van der Waals links, and thus we keep a structure strength at a high level (see above).

Drying is performed by blowing 0% humidity air along the top surface of the sample at a given flow rate (0.3 m/s). In some cases the dish walls are either covered with a thin layer of perfluorinated grease (Krytox GPL 205, DuPont) allowing the sample to glide.

The distribution of apparent water along the sample axis is measured by a 0.5 T $^1$H MRI spectrometer with a one-dimensional double spin-echo sequence (two first echoes of the so-called CPMG sequence [16] with exponential extrapolation to compensate spin-spin relaxation) (see details on the sequence and data analysis in Appendix A). Each measured value of this distribution corresponds to the total water amount in a thin cross-sectional layer of 75 μm thickness along the axis of the Petri dish. The total NMR signal in the sample is also measured separately, providing a reliable measure of the total water amount at each time down to very low values, in particular when NMR data become too noisy.

Finally, since only the presence of liquid water is detected on MRI profiles, our measurements cannot provide the effective sample thickness especially if some near free surface region is dry. We therefore place upon the central point of the sample a sealed tripod pot (3 cm diameter) containing a solution with a similar NMR behaviour; the sample to pot size ratio and the distance (3 cm) between the pot and the sample are sufficiently large to negligibly affect the drying characteristics; imaging the position of this pot provides the exact position of the sample free surface in time.

## III. RESULTS

The direct observation of the sample makes it possible to identify different stages in the drying process (see Figure 1). Initially the gel adheres to the glass surface of the dish walls (adhesive samples). The observation of the samples from their upper side during drying reveals a first period (Regime A) where the gel starts to fracture along its periphery, which detaches it from the sidewalls of the container, then, additional open transverse fractures develop (see Fig. 1b), with exact patterns varying with the drying rate [14]. In this regime the gel pieces newly formed significantly shrink (compare Figs. 1b and 1c) while gliding over the substrate. After a while this process visually stops, the sample scatters light (see Fig. 1c), the free surface is opalescent, and we then enter the second regime (Regime B).

The opalescence of the sample likely occurs at the beginning of desaturation because in this regime small patches of saturated region coexist with unsaturated regions around the free surface of the sample, at a length scale which scatters visible light (micrometer scale). Opalescence would thus disappear when further drying decreases the length scale of those patches down to the nanometric scale, so that our system reappears as homogeneous for visible light wavelength.

Shortly after, the pieces develop fractures (opening width of the order of 2 μm, see inset of Fig.3) without further significant shrinkage (see Fig.1d upper), an effect reminiscent of that observed for confined colloidal suspensions drying in channel [15]. None of these fractures fully cross a piece which remains as a continuous section. At the end of this phase the gels become translucent again.

By contrast, once the dish walls are covered with a thin layer of grease, the gel is able to freely glide on its substrate Fig1.a',b',c' and d'. We observe no fracture during the first regime, that we will also call Regime A: the sample simply shrinks (compare Figs. 1a', 1b', 1c'). There is again an opalescence at the transition (see Fig. 1c') to a regime B, with similar fractures nucleating without apparent further shrinkage (see Fig. 1d').

The basic difference in the evolution of adhesive and non-adhesive samples is during Regime A, the absence of open fractures followed by further shrinkage. The fractures observed in regime A in the adhesive case are due to tensile stresses resulting from two opposite tendencies: radial shrinkage due to water withdrawing and bounding of the bottom interface of the sample. On the contrary in regime B the second type of fractures develops in a non-shrinking situation, which means that there is no macroscopic deformation in the sample and thus no macroscopic tensile stress. This means that these fractures occur as a result of local high stresses in relation with air penetration in the sample. The validity of this scheme is now investigated though NMR which yields clear information on the water saturation ($\psi$, water to void volume ratio) in the material during the drying process.

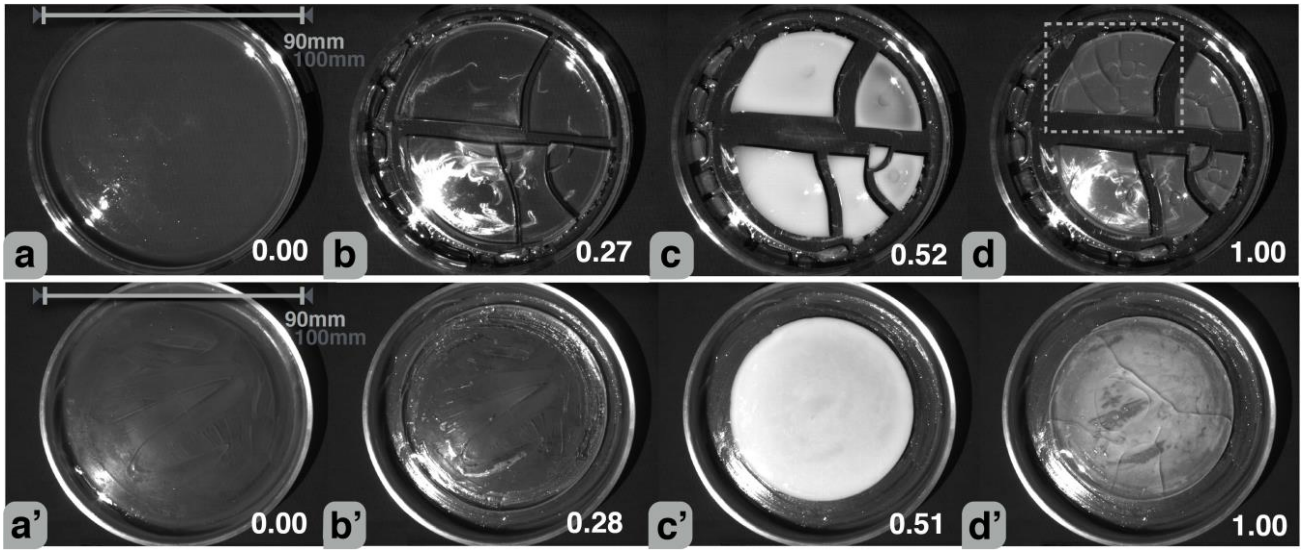

**Figure 1:** Aspects of drying gel layers ( $\phi_0 = 20\%$ ) on an adhesive (upper row) or non-adhesive (lower row) substrate at different stages of the process: (a) and (a') initial state; (b) $\phi = 39\%$ and (b') $\phi = 37\%$ Regime A (shrinkage associated (b) or not (b') with open fractures); (c) $\phi = 55\%$ and (c') $\phi = 51\%$ transition between regimes A and B; (d) and (d') same concentrations, Regime B (new fractures). The white numbers indicate the current time to total drying duration ratio.

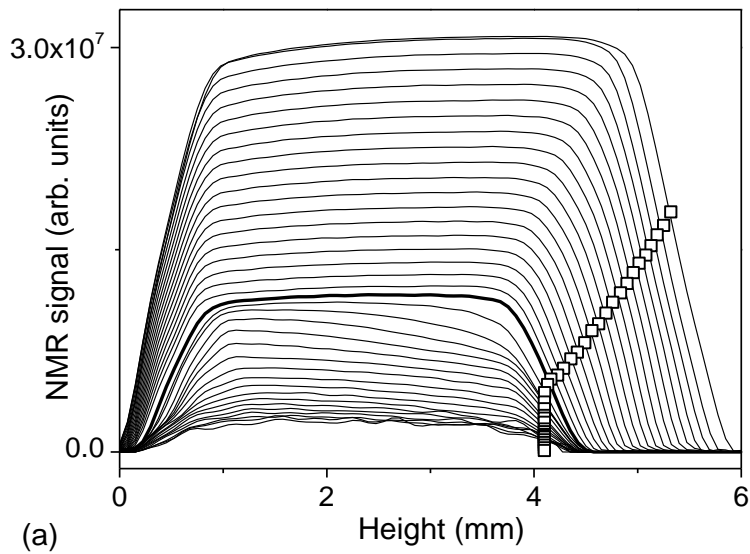

(a)

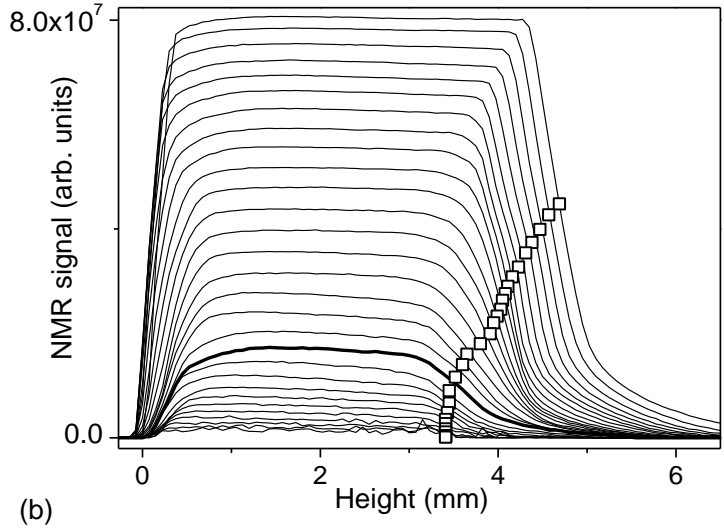

(b)

**Figure 2**: Distributions of water content in time inside a non-adhesive (a) and adhesive (b) gel ($\phi_0$ =20%) during drying. The continuous curves from top to bottom correspond to successive times (every 30 min. for (a) and every 55 min. for (b)) from the test beginning. The symbols are the corresponding positions of the sample free surface as measured (by NMR) from the position of the water pot. The thick line corresponds to the transition between the two regimes (see text).

The water content for adhesive and non-adhesive samples continuously decreases in time as shown by the decrease of the profile level (see Fig.2). Crucially, the position of the sample free surface exactly follows the position of the end of the plateau in the water profiles (see Fig.2) assuring that no significant dry region around the top of the sample exists. During a first period both the profile level and extent decrease, indicating a loss in water content in the radial and axial directions; furthermore, we observe the profiles holding parallel to each other, signifying water remains homogeneous distributed along a vertical axis. In a second regime (profiles starting from the thick line, see Fig.2) the sample thickness remains approximately constant while the profile level goes on decreasing.

Note the linear ramp shape at bottom and top sides of profiles at the beginning of the experiment, and the non-zero water amount above the measured position of the free surface. Both are due to the bending (see Appendix B) of the sample which becomes slightly concave (with an almost constant radius of curvature of about 1 m) rapidly after the test beginning [17], so that squares in Figure 2 only indicate some average position of tripod's legs on the curved sample surface. For the adhesive samples the measurements correspond to the average water amount through different material pieces. This suggests that each of the pieces resulting from fractures behave in a similar way, but here the analysis of the water content profiles is not straightforward since the different pieces formed with different sizes, shapes and curvatures may evolve with somewhat different timings.

## IV. DISCUSSION AND ANALYSIS

The critical question is whether the samples are saturated ($\psi = 1$) or not during desiccation. In order to clarify this we use our measure of the total water amount in the sample ($\Omega_{water}$), the sample thickness ($h$) in time as measured from NMR profiles, and the horizontal apparent area (neglecting the impact of bending) of the sample ($S$) measured from videos taken from above during the test, which provides the apparent volume $\Omega = Sh$. We deduce the saturation: $\psi = \Omega_{water}/(\Omega - \Omega_p)$, in which $\Omega_p = \phi_0 \Omega(t=0)$ is the volume of particles, and the solid concentration: $\phi = \Omega_p/\Omega$. The reproducibility of the MRI data is excellent (see Appendix C) with an uncertainty of less than 2.5%. The main uncertainties on the present data result from slight variations of sample shape and the uncertainty on the estimation of its section area, which explains some saturation values larger than 1 through our above computations.

From the evolution of $\psi$ as a function of $\phi$ two regimes clearly appear (see Fig. 3). In a first regime $\psi$ remains equal to 1, indicating that a negligible amount of air penetrates the gel, and the sample exhibits isotropic shrinking, i.e. its thickness to apparent radius ($R = \sqrt{S/\pi}$) ratio remains constant. The material finally reaches a solid concentration which is not exactly the maximum packing fraction in a disordered configuration for non-colloidal particles ($\approx 60\%$), but here the initial loose system was already an aggregated network. In a second regime $\phi$ first slightly increases [18] then remains constant while the saturation drops to zero. The results are the same for a sample fracturing in regime A. This means that we would get the same overall characteristics (shape and water content) for a non-adhesive and an adhesive sample by gathering the pieces formed in the latter case.

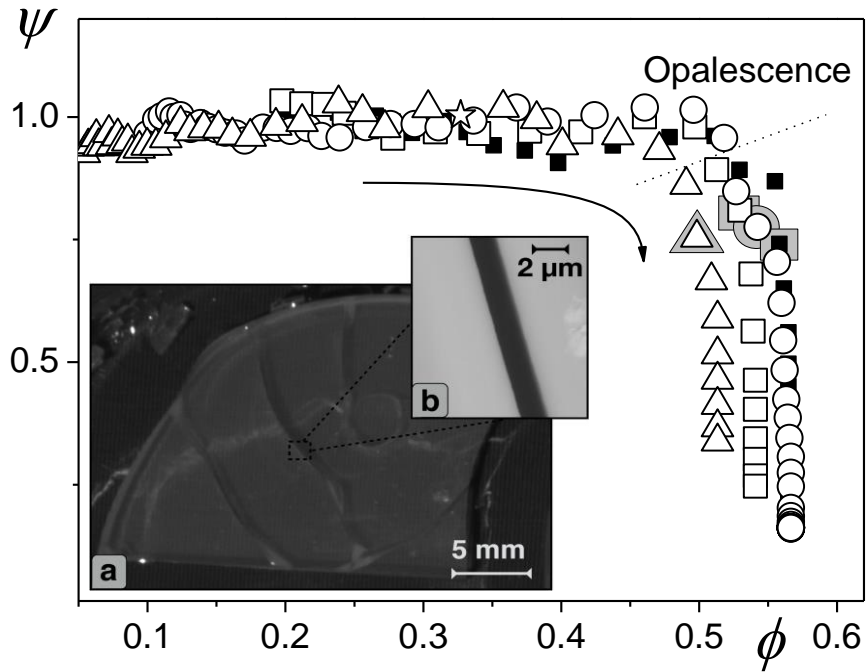

**Figure 3**: Saturation as a function of particle concentration for different $\phi_0$, $h_0$ and $V_e$: (non-adhesive) (squares) 20%, 5 mm, $0.33\,\text{mm.h}^{-1}$, (circles) 10%, 10 mm, $0.23\,\text{mm.h}^{-1}$ (triangles) 5%, 5 mm, $0.13\,\text{mm.h}^{-1}$; (adhesive): (filled squares) 20%, 5 mm, $0.21\,\text{mm.h}^{-1}$. Grey symbols show the time of appearance of the second type of fractures. The arrows show the drying progression. The inset shows typical aspect of a fractured piece (a) corresponding to the dotted square of Figure 1, and a fracture of type 2 inside (b). On each test the uncertainty on $\psi$ and $\phi$ is 10%.

Since in regime B the solid structure is fixed, we can deduce the saturation profiles through the sample ($\chi(x,t)$, where $x$ is the distance from sample bottom) (see Fig.4 top inset) from the water content profiles by scaling the NMR signal by that measured at the same position just before desaturation starts. After a short transient period the saturation profile keeps a constant shape, i.e. the profiles rescaled by the mean saturation at each time ($\psi(t) = \langle \chi(x,t) \rangle$) fall along a master curve (see Fig.4 bottom inset). This slightly inclined curve shows that the saturation gently decreases from the bottom to the free surface.

In such a case, most of the air volumes inside the sample are saturated by vapor [12, 19] so that evaporation mainly occurs around the air-liquid interfaces situated close to the sample free surface. Indeed a process of vapor diffusion from the interior of the sample cannot be significant since inside the sample the air volumes surrounded by liquid patches are saturated with vapor, which means that the vapor flux associated with such a diffusion process towards the sample free surface would be extremely slow. That also means that if the liquid was distributed in independent patches in the sample, the main vapor flux would come from the patches close to the free surface which would be rapidly dried. Since this process would start again at the next depth in the sample we would observe a drying front growing in depth. Our observation of the progressive homogeneous desaturation thus proves that there is a continuous liquid film coating the particles. Moreover if some dewetting process was possible somewhere it would also imply the possibility of a dewetting around the free surface of the sample thus leading to a growing dry front. As a consequence the only possibility is a continuous film coating all the particles.

This implies that the liquid forms a continuous film coating the particles, which flows through the bead packing and evaporates around the free surface. Here the original situation is that this flow involves a continuous liquid layer of very small thickness (of the order of a few molecule sizes) which extends from the bottom to the top of the sample. Indeed if we assume that each particle is covered with a liquid layer of uniform thickness $e$, for a particle concentration of 50% we find $e \approx \psi r/3$, which implies that for $\psi \approx 0.5$ we have $e \approx 1$ nm. In contrast, for larger particles, a dry front is usually observed after a first period of almost homogeneous desaturation (see for example [20]).

A liquid flow in such a nanoporous medium under the action of capillary effects would be extremely slow. The micron-sized cracks can only slightly help this flow. As soon as some water is removed from the free surface (over a thickness $\delta$) evaporation then essentially occurs by diffusion from the first liquid-air interface at a distance $\delta$ from the dry air flow, with a characteristic velocity $V_d = (1/\rho)(1-\phi)D\Delta n/\delta$, in which $\rho$ is the liquid density, $n$ the vapour density and $D$ the coefficient of diffusion of vapour in air. A receding dry front of length $\delta$ (increasing in time) should form so that $V_d$ balances $V_e$, the vapour volume flux imposed by the external conditions (effective air flow along the sample free surface depending on sample shape and size), i.e. the drying velocity at the end of regime A. With our values of $V_e$ this would lead to a value for $\delta$ of several millimetres, in contrast with our observations. This implies that for nano-colloids there is an effect which precludes the formation of a dry front. This is likely the disjoining pressure [21] which will tend to maintain a continuous film of liquid molecules at the particle surface even under evaporation. Under these drying conditions this effect finally precludes any further evaporation since the saturation tends to a finite value (see Fig.4) which varies between 8 and 20% depending on exact experimental conditions.

The saturation profiles in time may be used to get precise information on the liquid dynamics in the regime B. The similarity of the profiles in time implies that we can write $\chi(x,t) = \psi(t)f(x)$. Moreover the profiles obtained under different conditions are approximately similar when $x$ is rescaled by $h$ (see bottom inset of Fig.4), which implies that for our tests $f$ is a single function of $x/h$. We can compute the average flow velocity of the liquid through the sample:

$$V(x,t) = (1-\phi)\int_0^x (\partial \chi/\partial t)dx = (1-\phi)\psi'(t)hF(x/h) \qquad (1)$$

with $F(x/h) = \int_0^{x/h} f(y)dy$. The average velocity inside the liquid network along the sample axis writes

$$v = V/(1-\phi)\chi = (\psi'/\psi)hF(x/h)/f(x/h) \qquad (2)$$

In particular at the top free surface of the sample we have $v(h) = \alpha(\psi'/\psi)h$ where $\alpha = F(1)/f(1) \approx 1.5 \pm 0.1$. The variations of $\psi$ in time thus governs the velocity variations. We find that $\ln\psi$ decreases linearly in time (see Fig.4), which means that for a given test $\psi'/\psi$ is a constant. Thus the velocity profile $v(x)$ does not vary in time.

More precisely, for our different tests, for $\psi$ as a function of the time rescaled by the characteristic drying time ($T = h/V_e$), i.e. $\theta = t/T$, we find a single curve: $\log\psi \approx -\theta$. Finally we deduce the general result:

$$v(h) \approx (3.4 \pm 0.4)V_e \qquad (3)$$

which is valid as soon as an air path throughout the sample exists. Thus we have a liquid flow in a layer which thins in time but the average velocity in this layer is constant, independent of sample thickness and proportional to the external air flux. Under these conditions the evaporation is apparently governed by local phenomena at sample surface. This situation is completely different from that observed for usual porous media, with larger pores, for which the drying rate remains constant during most of the desaturation process as a result of fast capillary equilibration processes [12-13, 22]. For nano-particles there is an almost homogeneous desaturation process but the drying rate is governed by the thickness of the liquid layer covering the particles, and not by the thickness of the dry region (which apparently does not exist). Thus, in this situation it seems that when the liquid film over the particles is of nanometric size it becomes able to ensure a relatively easy draining of the liquid towards the free surface, which avoids the formation of a dry front. This suggests that, in that case, both the mobility of the liquid is higher than usual (i.e. smaller apparent viscosity) and its ability to evaporate is lower, but further study is needed to understand the origin of these effects.

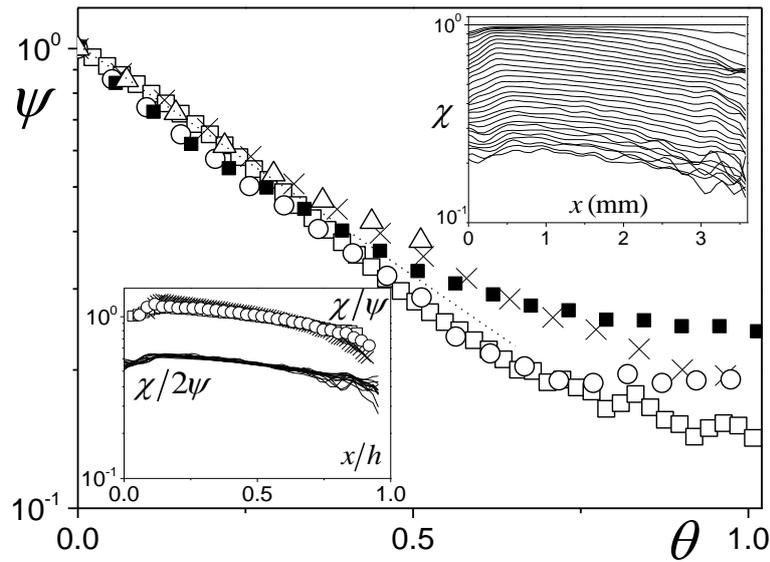

**Figure 4**: Average saturation as a function of dimensionless time (same symbols as in Fig.3, crosses correspond to $\phi_0 = 20\%$, $h_0 = 10\,\text{mm}$, $6.9\,\text{mm.h}^{-1}$); the dotted straight line is a guide for the eye. Top inset: saturation profiles (every 15 min.) in Regime B for $\phi_0 = 20\%$, $h_0 = 5\,\text{mm}$. Bottom inset: (lines) same saturation profiles as in top inset rescaled by twice the average saturation; (symbols) average rescaled saturation profiles for samples at 20 and 10%.

## V. CONCLUSION

Our results show that during most of the time in Regime B liquid coexists with significant air volumes and flows towards the sample free surface in layers which progressively thin down. Thus the internal stresses do not significantly evolve during this period. The situation differs from Regime B when air for the first time progressively penetrates the sample (see Fig.4 top inset). High local stresses, induced by disjoining pressure effects when air fingers enter the thin liquid films, can push close particles against each other, and tend to further concentrate the system which is already well packed. Only some slight rearrangements extending over large distances are possible, leading to failures then allowing slight further concentrations in separate regions. Thin fractures will thus occur in the sample during the short transient period at the beginning of Regime B. This conclusion is consistent with our observations of the approximate time of appearance of these fractures (see Fig.3).

These different observations show how the air penetrates through a porous medium and its impact on fractures. The direct analysis of the water dynamics at the molecular scale through a compacted sample show that liquid flows at the nanoscale have unexpected specific properties.

**Acknowledgments**: This paper benefited from useful discussions with G. Scherer, E. Dufresne and F. Boulogne. We are grateful to David Hautemayou for his technical support. We warmly thank Max Zieringer et Adrian Pegoraro for their useful advices.

**Appendix A. MRI technique**

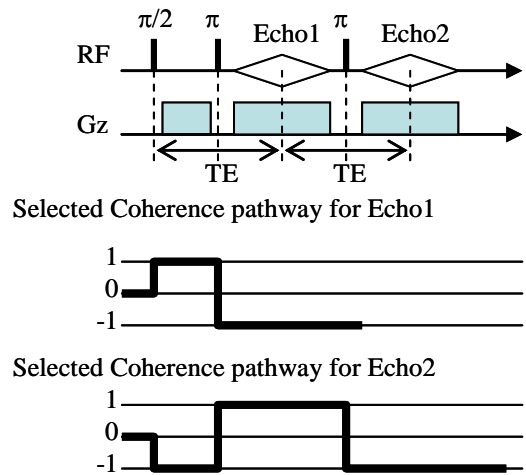

**Figure A.1**: (Color online) NMR sequence used in our experiments.

The NMR profiling sequence makes use of pulsed field gradients, as shown in Figure A.1. Both coherence pathways are selected by means of a 8 steps duplex cogwheel phase cycling scheme acting on $\pi$ pulses only. The gradient strength is $G_z = 0.04\,\text{T}.\text{m}^{-1}$, which is actually close to the maximum gradient strength available on our hardware, with $500\,\mu\text{s}$ stabilizing time, and freshly tuned pre-emphasis which removes eddies effects at the end of this stabilizing time. Our echo time (which also more or less corresponds to the time length of each echo recording) was $\text{TE} = 8.4\,\text{ms}$ which, according to Shannon rules provides the theoretical space resolution:

$$\delta x_{th} = \frac{2\pi}{\text{TE}} \frac{1}{\gamma G_z} \approx 70\,\mu\text{m}$$

The experiment was performed at the magnetic center of our gradient coil (a BGA26 by Bruker, 26cm inner diameter), which guaranties minimal profile distortion due to possible gradient non linearity (at least well under our space resolution). Finally, prior to the experiment, our magnetic field was shimmed so as to get a $\Delta f = 50\,\text{Hz}$ large spectral line for our sample. The blurring to be expected on our NMR profiles due to field inhomogeneities is then:

$$\delta x_{blur} = \frac{2\pi \Delta f}{\gamma G_z} \approx 30\,\mu\text{m}$$

The effective space resolution on our data was then estimated as:

$$\delta x_{eff} = \sqrt{\delta x_{th}^2 + \delta x_{blur}^2} \approx 76\,\mu\text{m}$$

Our measurements were not subjected to any $T_1$ weighting, because the recycling delay between sequences was set to 5 times the spin-lattice relaxation time in our sample at the beginning of the experiment (and this time was not prone to increase during sample drying).

In order to remove the spurious contribution of spin-spin relaxation, we proceeded as follows. Assuming a mono-exponential relaxation in each layer of our sample, with a local relaxation time $T_2(z)$, and denoting $m_0(z)$ the actual spin density profile, the profile obtained from first and second echoes were modelled respectively as:

$$\text{prof}_1(z) = m_0(z) \exp\left(-\frac{\text{TE}}{T_2(z)}\right) \quad \text{and} \quad \text{prof}_2(z) = m_0(z) \exp\left(-\frac{2\text{TE}}{T_2(z)}\right)$$

The profiles presented in the article were then calculated following the profile extrapolation formula:

$$\text{prof}(z) = \frac{\text{prof}_1(z)^2}{\text{prof}_2(z)} = m_0(z)$$

which is supposed to remove the space-dependent $T_2(z)$ contribution. Note that this formula still provides a fair quantitative correction even if $T_2$ exhibits slight fluctuations inside each sample slice, and may be shown to correct for imperfections (if any) of $\pi$ pulses.

The correctness of the (quasi-) exponential assumption for signal decay in each single pixel was checked experimentally. Using a sequence with a larger number of echoes, the evolution of NMR signal intensity for echo times ranging between 8 and 120 ms was followed for a few pixels, in the case of a partially desaturated sample which seems to us the most critical case. It was seen that for any given pixel, recorded data can be accurately fitted by a decreasing exponential function, within the noise level. It was also seen that the relaxation rate strongly evolves from one pixel to another (depending on local water amount), thus justifying the use of a position-dependent relaxation time in our theory. At last, the smallest observed value of $T_2$ value at the end of a drying experiment was about 10ms, which is still in the proper range - regarding our echo times - for a proper use of our extrapolation formula.

**Appendix B. Impact of sample curvature**

The impact of sample curvature on the NMR profiles is illustrated in Figure B.1 in the case of a spherical shape of the sample, with a large radius of curvature. In that case a simple mathematical integration of the total water volume in each cross-section shows that the resulting gradient in water content observed at the top and bottom of the sample is a constant, giving a straight inclined line at the beginning and end of the profile. In reality the sample shape is not exactly spherical, which explains a slight difference with regards to a straight line.

The effective sample shape evolution can depend on the rate of drying, as shown for the test of Figure B.2 which corresponds to the same conditions as in Figure 2a except that the air flow velocity was much lower. In that case there is a first stage during which the sample significantly bends but in a second stage its curvature tends to cancel. Despite these differences the evolution of the water content profiles are similar to those observed for a larger air flow velocity: we observe the same two regimes and a transition at almost the same solid volume fraction.

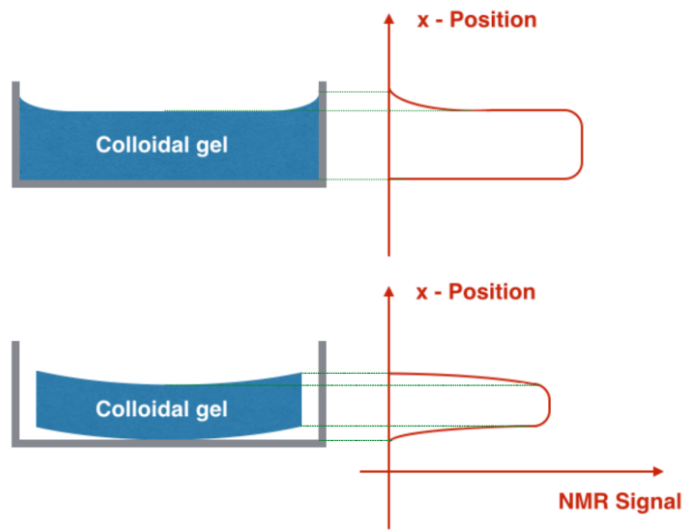

**Figure B.1**: (Color online) Scheme of sample shape in a radial cross-section and corresponding NMR profiles of the water content for a fully saturated sample at the beginning (upper view) or after some contraction (lower view).

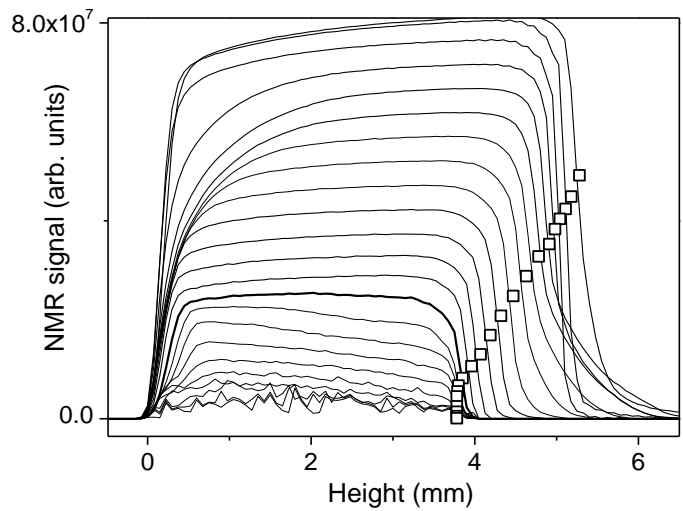

**Figure B.2:** Water content profiles in time for a test under the same conditions as in Figure S.2 but a lower air flow velocity. The curves correspond to successive times (every 55 min.) and the squared symbols correspond to the sample free surface similarly as in Figure 2a.

**Appendix C. Reproducibility of MRI data**

The uncertainty on data mentioned in the text is extracted from identical tests. We observe that the set of water content profiles for two identical experiments evolve very close to each other all along the drying process; the two same drying regimes and position of the transition superimpose properly (see Figure C.1). This good correlation with time and water content enables us to validate a good control of the external conditions. Even if we acknowledge a difference in the effective drying rate for the 2 samples (see inset of Figure C.2), it is very light and could be due to slight differences in the sample surface shape originating from the initial state or during drying therefore leading to slight variations of the characteristics of the air flow along the sample surface. As the deformation (curvature) of the sample can be slightly different (maximum difference: 2%), this may also lead to slightly different shapes of the saturation profiles around the edges too. However despite these variations the uncertainty on the saturation and solid fraction remains very low, i.e. 2.5% (maximum relative difference between corresponding values on each profile).

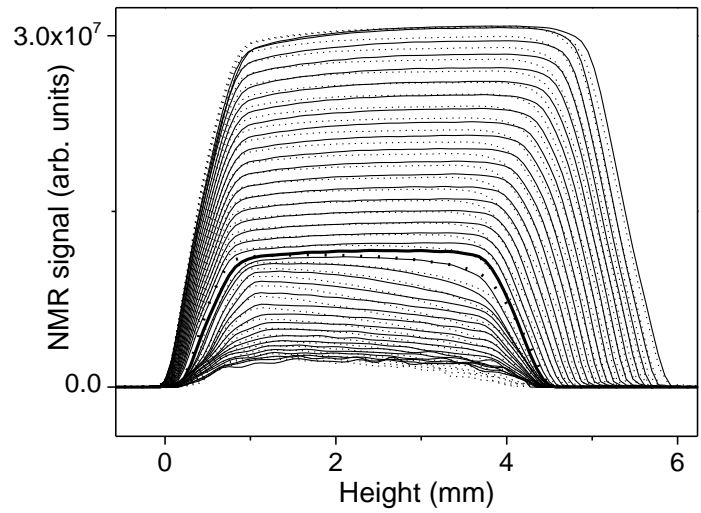

**Figure C.1**: Water content profiles in time for two identical tests under conditions of Figure 2a. Data of Figure 2a correspond to continuous lines, the other data to dotted lines.

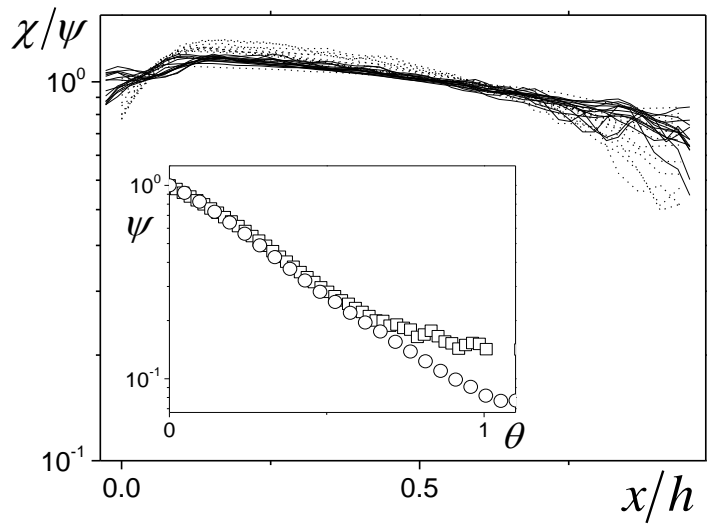

**Figure C.2**: Saturation profiles in Regime B for the two sets of data of Figure C.1 rescaled by the average saturation. The inset shows the average saturation as a function of dimensionless time for these two tests (data of Figure 2a as squares, the other data as circles).